\documentclass[twocolumn,prb,floatfix,showpacs]{revtex4}
\usepackage{graphicx}

\newcommand{\be}{\begin{equation}}
\newcommand{\ee}{\end{equation}}
\newcommand{\barr}{\begin{eqnarray}}
\newcommand{\earr}{\end{eqnarray}}

\begin{document}

\title{Optical bistability in subwavelength apertures containing
nonlinear media}

\author{J. A. Porto,$^1$ L. Mart\'{\i}n-Moreno,$^2$ and F. J. Garc\'{\i}a-Vidal,$^1$}

\affiliation{
$^1$Departamento de F\'{\i}sica Te\'orica de la Materia Condensada, Facultad de Ciencias
(C-V), Universidad Aut\'onoma de Madrid, E-28049 Madrid, Spain \\
$^2$Departamento de F\'{\i}sica de la Materia Condensada, ICMA-CSIC,
Universidad de Zaragoza, E-50015 Zaragoza, Spain}

\date{\today}

\begin{abstract}
We develop a self-consistent method to study the optical response
of metallic gratings with nonlinear media embedded within their
subwavelength slits. An optical Kerr nonlinearity is considered.
Due to the large E-fields associated with the excitation of the
transmission resonances appearing in this type of structures,
moderate incoming fluxes result in drastic changes in the
transmission spectra. Importantly, optical bistability is obtained
for certain ranges of both flux and wavelength.
\end{abstract}
\pacs{78.20.Ci,42.65.Pc,73.20.Mf}

\maketitle



Since the appearance of the photonic crystal concept, there has
been a strong interest in the optical properties of nano- and
micro- structured systems. This is due, in part, to the potential
applications for small all-optical devices. An interesting
possibility is to include some non-linear elements in the
structures \cite{john93,sca94,fle03}, which may result in optical
switches and gates. Systems based on nonlinear photonic crystals
presenting optical bistability have been recently proposed
\cite{cen00,sol02,sol03,yan03}. Another promising route for
nonlinear optics is to use structured metal films. The enhanced
optical transmission phenomenon found in subwavelength hole
arrays\cite{ebb98} and the high local field enhancements
associated with it\cite{krishnan}, suggest the possibility of
strong non-linear transmission effects if nonlinear media are
embedded in the structure. Recently, photon transmission gated by
light of a different wavelength were reported \cite{smo02}, and
the possibility of bistability in enhanced transmission has been
mentioned \cite{dyk03}. However, up to our knowledge, a detailed
calculation for such metallic structures containing nonlinear
media had not yet been done.

In this paper, we present a theoretical analysis of the
transmission properties of a one-dimensional metallic grating with
subwavelength slits \cite{sch98,por99,wen00,col01,bar02}, filled
with a Kerr nonlinear media. As we will discuss below, even this
simple system presents interesting optical properties, as optical
bistability for certain ranges of wavelength and incident flux.

\begin{figure}[tb]
\includegraphics[angle=0, width=8cm]{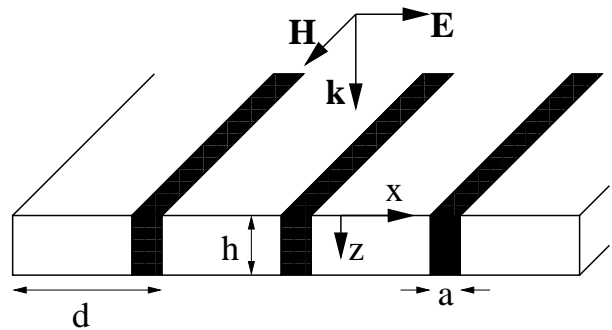}
\caption{Schematic view of the metallic structure under study: a
metallic grating of period $d$, film thickness $h$, and slit width
$a$. Inside the slits, we consider a Kerr nonlinear material,
characterized by a Kerr coefficient $n_2$. Light is considered to
impinge at normal incidence with an E-field pointing across the
slits.}
\label{fig1}
\end{figure}

In Fig.~\ref{fig1} we show a schematic view of the structure under
study. The chosen parameters of the grating are: $d=0.75$~$\mu$m,
$a=0.05$~$\mu$m, and $h=0.45$~$\mu$m. This set of parameters is
typical for experimental studies of subwavelength apertures in the
optical regime \cite{Lezec02}. The slits are supposed to be filled
with a Kerr nonlinear media, whose dielectric constant at point
$\vec{r}$ depends on the intensity of the E-field at this point,
$|E(\vec{r})|^2$,

\begin{equation} \epsilon(\vec{r}) = \epsilon_l +
\chi^{(3)}|E(\vec{r})|^2 \label{kerr}
\end{equation}

where $\epsilon_l$ is the value of the dielectric constant at low
intensity, and $\chi^{(3)}$ is the third order susceptibility
related to the Kerr effect. Since the change of the dielectric
constant is not very large, it is common to approximate the
refraction index as: $n(I)= n(0) + n_2 I$, where $I$ is the
intensity of the optical field (measured in units of energy flux),
and $n_2$ is known as the Kerr coefficient. This allows to express
the incident flux, $I_0$, in units of $n_{2}^{-1}$ and the results
will be valid for all Kerr materials with the same $n(0)$. In what
follows we consider $n(0)=$~1.5 ($\epsilon_l$=2.25).

The electromagnetic (EM) properties of these gratings are analyzed
by a modal expansion of the electric and magnetic fields. Two
simplifications are incorporated to the exact modal expansion: we
consider perfect metal boundary conditions and, as the slit width
is much smaller than the wavelength, only the fundamental
eigenmode in the modal expansion of the EM fields inside the slits
is included. We have demonstrated in previous theoretical works
\cite{gar02} that these are reasonable approximations when
analyzing optical properties of nanostructured good metals like
silver and gold. Moreover, within this framework, we have been
able to reproduce in semi-quantitative terms the phenomena of
beaming \cite{LMM03} and enhanced transmission \cite{gar03},
appearing for subwavelength apertures in the optical regime. We
consider p-polarized light (E-field pointing across the slits),
for which both extraordinary transmission and electric field
enhancements have been reported. Furthermore, in this paper we
concentrate on light impinging at normal incidence onto the
structure.

In order to account for the nonlinear response of the material
within the slits, an iterative self-consistent method is used. For
this, the slit region is divided into $N$ narrow slices,
perpendicular to the $z$-direction. For a given iteration, the
magnetic field in the n-th slice is expressed as
\begin{equation}
H_y = \frac{1}{\sqrt{a}}\left\{A_n e^{\imath\sqrt{\epsilon_n}k_0z}
+ B_n e^{-\imath\sqrt{\epsilon_n}k_0z}\right\} \label{exp}
\end{equation}
where $k_0$ is the wavenumber of the incident light (in vaccuum),
$\epsilon_n$ is the dielectric constant in the n-th slice, and
$A_n$ and $B_n$ are the modal expansion coefficients. Within each
slice, the electric field can be obtained from
$E_x=(-\imath/\omega\epsilon\epsilon_0)(\partial H_y/\partial z)$.
By matching $E_x$ and $H_y$, the coefficients of the modal
expansion in two consecutive strips can be related as follows,

\begin{equation}
\left( \begin{array}{c} A_{n+1} \\ B_{n+1} \end{array} \right) =
\frac{1}{2} \left( \begin{array}{cc} C_+ \Psi_{n,-}  &
C_- \Psi^{-1}_{n,+} \\
C_- \Psi_{n,+}  & C_+ \Psi^{-1}_{n,-} \end{array} \right) \left(
\begin{array}{c} A_{n} \\ B_{n} \end{array} \right) \label{mat}
\end{equation}

with the definitions
$C_{\pm}=1\pm\sqrt{\epsilon_{n+1}/\epsilon_{n}}$,
$\Psi_{n,\pm}=\exp\left[\imath(\sqrt{\epsilon_{n}}
\pm\sqrt{\epsilon_{n+1}})k_0z_n\right]$, and $z_n$ marking the
position of the interface between slices $n$ and $n+1$. The modal
coefficients at the entrance and exit of the slit are related
through the product of the matrices corresponding to the matching
of EM fields between consecutive slices. Finally, the connection
with the incoming and outgoing waves outside the grating is
established, obtaining the transmission and reflection
coefficients for the whole structure. In the next step of the
iterative procedure, a new set of $\epsilon_n$ is obtained through
Eq.(1) from the previous calculation of the E-fields in the
slices. For a given value of $N$ and $I_0$, this linear response
calculation is repeated until convergency in the set of
$\epsilon_n$ is reached. And then, for each value of $I_0$, $N$ is
increased until the transmitted flux does not depend on $N$.
Still, the described procedure does not completely specify the
solution: this non-linear system may present a multiplicity of
solutions, and the transmitted flux obtained at the end may depend
on the profile of $\epsilon(z)$ chosen as seed of the iterative
procedure. Two choices are used in this paper: for a given
incident flux we use as seeds of the iterative procedure the
converged dielectric constant profiles obtained from a previous
calculation for either smaller or larger incident fluxes.

\begin{figure}[tb]
\includegraphics[angle=0, width=8.5cm]{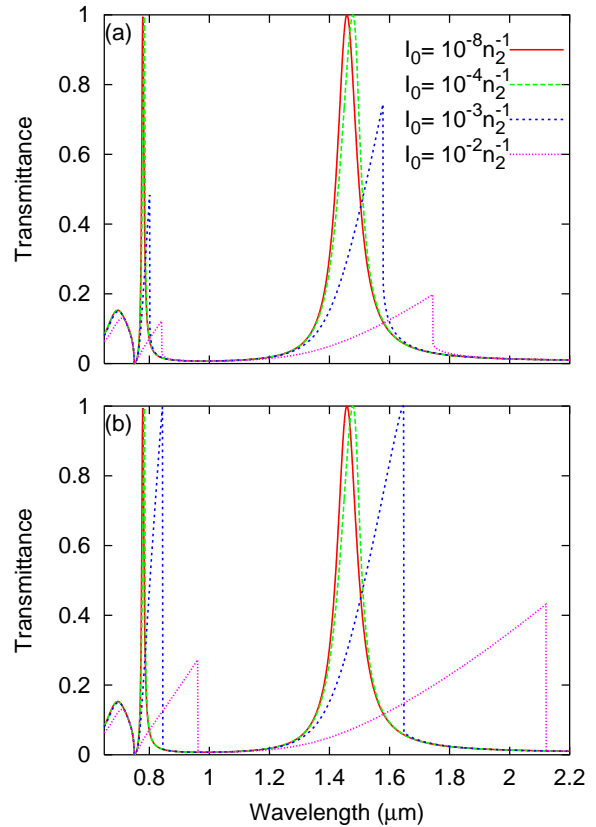}
\caption{ Transmittance versus wavelength for different values of
incident flux obtained while increasing (a) and decreasing (b) the
incident flux. The parameters of the metallic gratings are:
$d=0.75\mu$m, $a=0.05\mu$m, and $h=0.45\mu$m. The incident fluxes
considered are (1) $10^{-8}n^{-1}_2$, (2) $10^{-4}n^{-1}_2$, (3)
$10^{-3}n^{-1}_2$, and (4) $10^{-2}n^{-1}_2$.} \label{fig2}
\end{figure}

Fig.~\ref{fig2}(a) renders the transmittance (total transmitted
flux divided by incident flux) versus wavelength, for different
values of $I_0$ in units of $n_{2}^{-1}$. For the chosen
parameters, in the linear regime (red curves in both panels (a)
and (b)), the metallic grating presents two transmission
resonances, appearing at $\lambda=1.46 \, \mu$m and $\lambda=0.78
\, \mu$m, very close to the period of the grating. These two peaks
are representative of the two types of transmission resonances
appearing in metallic gratings \cite{por99,col01}: i) slit
waveguide modes in which the EM-fields are highly concentrated
within the slits and ii) coupled surface plasmon polaritons (SPP)
excited at the entrance and exit interfaces of the structure. In
Fig.~\ref{fig2}(a), the transmittance at a given wavelength has
been obtained by considering the converged situation at a lower
incident flux (for that particular wavelength) as the seed for the
iterative scheme. As can be seen, as the incident flux is
increased, the peaks shift to larger wavelengths and their
magnitude decreases. In Fig.~\ref{fig2}(b) we represent the same
quantity, transmittance versus wavelength, for the same structure,
{\it but for decreasing incident fluxes}. For a given wavelength,
the flux is first increased up to $I_{\mathrm{max}}=0.1
n_{2}^{-1}$, which is a representative value of a ``high-flux''
situation, and then decreased slowly to the value $I_0$, using as
seed for the iterative procedure the converged solution obtained
for a higher flux. Remarkably, the transmission spectra differ for
increasing and decreasing fluxes, which is a clear manifestation
of {\bf optical bistability}. Besides, for decreasing incident
flux, the transmission spectra depends on the initial incident
flux $I_{\mathrm{max}}$. In particular, in Fig.~\ref{fig2}(b) the
transmittance peaks for incident flux 10$^{-2} n_2^{-1}$ would
increase (and even reach unity) for larger values of
$I_{\mathrm{max}}$.

\begin{figure}[tb]
\includegraphics[angle=0, width=8cm]{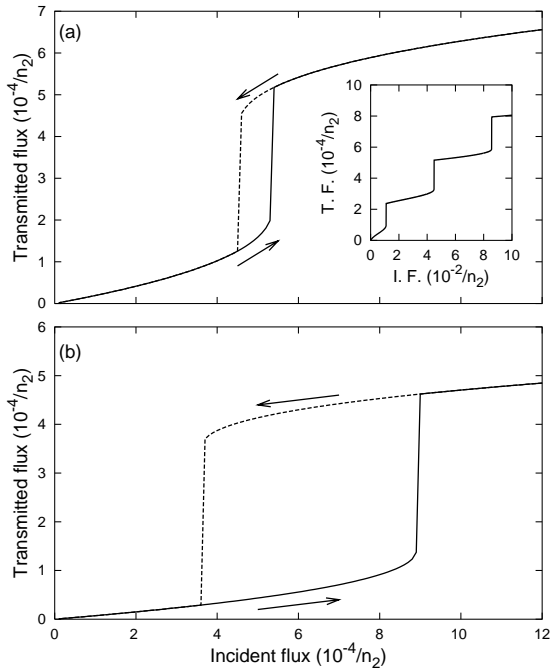}
\caption{ Transmitted flux versus incident flux for wavelength (a)
$\lambda=1.55\mu m$ and (b) $\lambda=0.8\mu m$ for a metallic
grating with parameters: $d=0.75\mu$m, $a=0.05\mu$m, and
$h=0.45\mu$m. Solid (dashed) lines correspond to increasing
(decreasing) fluxes. In the inset of panel (a) it is shown the
transmitted flux versus increasing incident flux for
$\lambda=1.55\mu m$ for a metallic grating of the same $a$ and $d$
than in the panels but with $h=13\mu$m.} \label{fig3}
\end{figure}

Optical bistability is better analyzed considering the situation
for a fixed wavelength. In Fig.~\ref{fig3}(a), we show the
transmitted flux as a function of $I_0$ for $\lambda=1.55\mu m$,
which is the typical wavelength used in telecommunications (and
slightly larger than the wavelength of the waveguide mode in the
structure under consideration in this paper). A clear bistable
loop is observed at a range of incident fluxes.
 When increasing the
incident flux (full line in Fig.~\ref{fig3}(a)), the transmitted
flux increases. At a certain incident flux (for this particular
case around $5.4 \times 10^{-4}/n_2$), the transmitted flux jumps
to a higher value in a discontinuous manner. By contrast, the
system has a different behavior when the incident flux is
decreased. The system, now in a high transmittance situation,
instead of jumping back to low transmitted flux, follows the upper
branch in Fig.~\ref{fig3}(a). In this system, where losses are not
included, the upper branch finishes when the ratio between the
transmitted flux and the incident flux reaches the unity. Then,
the transmitted flux jumps discontinuously to a lower value. The
presence of a narrow bistability region such as the one
illustrated in Fig.~\ref{fig3}(a) is an interesting situation for
device applications, as small changes in the incident flux would
imply significant changes in the transmitted flux.

Optical bistability is also obtained associated with the
transmission resonance linked to the excitation of two coupled
SPPs. In Fig.~\ref{fig3}(b), we show the bistable loop obtained
for a wavelength of $\lambda=0.8\mu$m, just slightly larger than
the period of the grating. The bistable region is wider than in
the previous case, mainly due to the higher incident flux needed
to jump to the upper branch when increasing the incident flux.

It is interesting to note here the similarities between the curves
rendered in Fig. 3 for increasing $I_0$ and the measurements of
light tunneling through individual pinholes in a granular gold
film covered with a layer of a nonlinear material
(polydiacetylene) reported in Ref. 21. These results were
explained in terms of a ``photon blockade'' phenomenon. In our
system, if the thickness of the metal film is large enough, the
spectral separation between different slit waveguide modes is
strongly reduced. Then, as the incident flux is increased, several
resonances can participate in the transmission process resulting
in a staircase behaviour of the transmitted flux versus the
incident one, as can be seen in the inset of Fig. 3a for a
metallic grating of $h=13 \mu$m. This behaviour is a clear signal
of the existence of ``photon blockade'' phenomena in the system
analyzed here. We believe that the structure proposed in this
paper (array of subwavelength apertures) is an alternative that
could allow the investigation of this interesting phenomenon in a
more controlled way.

We find that there are two characteristic EM-field patterns
associated, respectively, with the lower and the upper branch of
the bistability loop. At a given wavelength, the field pattern
within the slits for the lower branch corresponds approximately to
the field pattern obtained in the linear response calculation.
Even in this case, the lower branch presents, close to its
instability, non-linearities. However, these non-linearities are
associated more to the change in the {\it average} dielectric
constant inside the slit (via Eq.(~\ref{kerr})) than to changes in
the EM-field profile. The upper branch presents a different field
pattern which is, essentially, the one associated to the closest
transmission resonance located at a shorter
wavelength\cite{note_smaller}.

\begin{figure}[t]
\includegraphics[angle=0, width=8.5cm]{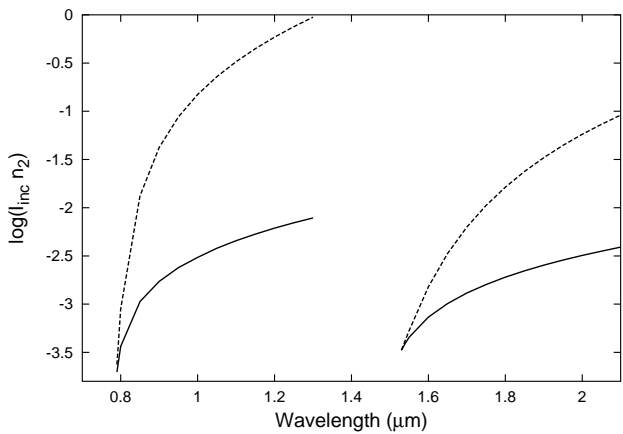}
\caption{Incident flux, in logarithmic scale, corresponding to the
beginning (continuous line) and end (dashed line) of the first
bistable region as a function of the wavelength for a metallic
grating with parameters: $d=0.75\mu$m, $a=0.05\mu$m, and
$h=0.45\mu$m.}
\label{fig4}
\end{figure}

Let us now study how the width of the bistability region depends
on the wavelength. In Fig.~\ref{fig4} we show the incident flux
corresponding to the start and the end of this region as a
function of the wavelength. For wavelengths in the interval
between $0.79$~$\mu$m and $1.53$~$\mu$m, bistability is related to
the shift to larger wavelengths of the coupled SPP resonance. The
width of the bistability region increases dramatically with the
wavelength. We arbitrarily terminate the curves at
$\lambda$=1.3$\mu$m, since the incident flux needed to jump to the
upper branch, which corresponds to the end of the bistability
region, becomes so high that our model is not valid any longer.
For wavelengths larger than $1.53$~$\mu$m, the first bistability
region is related to the shift to larger wavelengths of the slit
waveguide mode appearing approximately at $1.46$~$\mu$m at low
incident flux. Notice that the bistability regions do not start
exactly at the position of the linear-response transmission
resonances. As an approximate phenomenological rule we have found
that bistability regions start at slightly larger wavelengths, for
which transmittance is of the order of $1/4$ of the transmission
peak.

Given available fluxes and Kerr nonlinear media, optical
bistability should be experimentally accessible already. For
instance, some polymers can present a Kerr coefficient up to $10^{-8}
- 10^{-9} cm^2/W$. These values would imply incident fluxes of the
order of $10^4 - 10^5$ W/cm$^2$ for observing bistability, for the
parameters chosen in this article. For higher ratios between the
lattice parameter and the slit, given that the energy flux inside
the slits increases as $d/a$, bistability should occur for even
smaller values of the incident flux.

In summary, we have analyzed the optical response of a metallic
grating with subwavelength slits containing Kerr nonlinear media
by means of a self-consistent method. The nonlinear response
induces changes in the transmission spectra, with shifts of the
transmission peaks to larger wavelengths. In addition,
transmission spectra differ for increasing and decreasing
intensity. An important result is the observation of optical
bistability for certain ranges of wavelength and incident flux.
The levels of optical power needed for observing these effects are
achievable for typical Kerr nonlinear materials.

JAP gratefully acknowledges financial support from the Ram\'on y
Cajal Program from the Ministerio de Ciencia y Tecnolog\'{\i}a of
Spain. Financial support by the Spanish MCyT under contracts
MAT2002-01534 and MAT2002-00139 and by EU-STREP project ``Surface
Plasmon Photonics'' is gratefully acknowledged.




\end{document}